\documentclass[aps,prl,twocolumn,floatfix,superscriptaddress,showpacs,amsfonts,amssymb,amsmath,preprintnumbers]{revtex4}
\usepackage{bm}
\usepackage{graphicx}

\renewcommand{\eqref}[1]{Eq.\ (\ref{#1})}

\newcommand{\eqsand}[2]{Eqs.\ (\ref{#1}) and (\ref{#2})}
\newcommand{\Eqref}[1]{Eq.\ (\ref{#1})}

\newcommand{\Eqsand}[2]{Eqs.\ (\ref{#1}) and (\ref{#2})}
\newcommand{\figref}[1]{Fig.\ \ref{#1}}

\newcommand{\bea}{\begin{eqnarray}}
\newcommand{\eea}{\end{eqnarray}}
\newcommand{\beq}{\begin{equation}}
\newcommand{\eeq}{\end{equation}}
\newcommand{\lt}{\left}
\newcommand{\rt}{\right}

\newcommand{\dd}{\partial}
\newcommand{\vdel}{\boldsymbol{\nabla}}
\newcommand{\dpar}{\nabla_\parallel}

\newcommand{\vx}{\mathbf{r}}
\newcommand{\vv}{\mathbf{v}}
\newcommand{\vperp}{v_\perp} 
\newcommand{\vpar}{v_\parallel} 
\newcommand{\vu}{\mathbf{u}}
\newcommand{\du}{\delta u}
\newcommand{\dvu}{\delta \vu}
\newcommand{\dvuperp}{\dvu_\perp}
\newcommand{\duperp}{\du_\perp}

\newcommand{\vB}{\mathbf{B}}
\newcommand{\vb}{\hat{\mathbf{b}}}
\newcommand{\dB}{\delta B}
\newcommand{\dvB}{\delta\vB}
\newcommand{\dvBperp}{\dvB_\perp}
\newcommand{\dBperp}{\dB_\perp}
\newcommand{\dBpar}{\dB_\parallel}
\newcommand{\Epar}{E_\parallel}
\newcommand{\df}{\delta f}

\newcommand{\const}{\mathrm{const}}

\renewcommand{\Re}{\mathrm{Re}}

\newcommand{\Ma}{M}
\newcommand{\eps}{\epsilon}

\newcommand{\go}{\gamma_0}
\newcommand{\nuii}{\nu_{ii}}
\newcommand{\mfp}{\lambda_\text{mfp}}
\newcommand{\vthi}{v_{\text{th}i}}
\newcommand{\vthe}{v_{\text{th}e}}
\newcommand{\lvisc}{l_\nu} 
\newcommand{\lB}{l_B} 
\newcommand{\lRR}{L_{\rm RR}}

\newcommand{\kpar}{k_\parallel} 

\newcommand{\pperp}{p_\perp} 
\newcommand{\ppar}{p_\parallel} 

\newcommand{\xitr}{\xi_{\rm tr}}

\begin{document}

\preprint{Phys.\ Rev.\ Lett.~{\bf 100}, 081301 (2008) [e-print {\tt arXiv:0709.3828}]}

\title{Nonlinear Growth of Firehose and Mirror Fluctuations in Astrophysical Plasmas}
\author{A.\ A.\ Schekochihin}
\email{a.schekochihin@imperial.ac.uk}
\affiliation{Plasma Physics, Blackett Laboratory, Imperial College, %London, Prince Consort Road, 
London SW7 2AZ, United Kingdom}
\author{S.\ C.\ Cowley}
\affiliation{Plasma Physics, Blackett Laboratory, Imperial College, %London, Prince Consort Road, 
London SW7 2AZ, United Kingdom}
\affiliation{Department of Physics and Astronomy, UCLA, Los Angeles, California 90095-1547, USA}
\author{R.\ M.\ Kulsrud}
\affiliation{Princeton University Observatory, Princeton, New Jersey 08544, USA} 
\author{M.\ S.\ Rosin}
\affiliation{DAMTP, University of Cambridge, %Wilberforce Road, 
Cambridge CB3 0WA, United Kingdom}
\author{T.\ Heinemann}
\affiliation{DAMTP, University of Cambridge, %Wilberforce Road, 
Cambridge CB3 0WA, United Kingdom}
\date{\today}

\begin{abstract}
In turbulent high-beta astrophysical plasmas 
(exemplified by the galaxy cluster plasmas), pressure-anisotropy-driven 
firehose and mirror fluctuations grow nonlinearly to large amplitudes, 
$\delta B/B\sim1$, on a timescale comparable to the turnover time 
of the turbulent motions. The principle of their nonlinear 
evolution is to generate secularly growing small-scale magnetic fluctuations 
that on average cancel the temporal change in the 
large-scale magnetic field responsible for the pressure anisotropies. 
The presence of small-scale magnetic fluctuations may dramatically affect 
the transport properties and, thereby, the large-scale dynamics 
of the high-beta astrophysical plasmas. 
\end{abstract}


\maketitle

{\em Introduction.}---Many astrophysical plasmas are magnetized and weakly collisional, 
i.e., the cyclotron frequency $\Omega_i$ 
is much larger than the collision frequency $\nuii$ and the 
Larmor radius $\rho_i$ is smaller than the mean free path $\mfp$. 
In such plasmas, all transport properties, 
most importantly the viscosity and thermal conductivity, become 
anisotropic with respect to the local direction of the magnetic 
field \cite{Braginskii} --- even if the field is dynamically weak.\\ 
%This is true even in magnetic fields that are dynamically 
%weak, which means that while not affecting 
%the plasma motion via the Lorentz force, the field still exerts a powerful 
%influence by anisotropizing transport.\\
\indent
As a typical example where just such a physical situation is present 
is galaxy clusters \cite{SCKHS_brag,SC_dpp05}. While parameters 
vary significantly both within each cluster and between  
clusters, the weakly collisional magnetized nature of 
the intracluster medium (ICM) is well illustrated by 
the core of the Hydra A cluster, where 
$\Omega_i\sim10^{-2}$~s$^{-1}$, $\nuii\sim10^{-12}$~s$^{-1}$ and 
$\rho_i\sim10^5$~km, $\mfp\sim10^{15}$~km \cite{Ensslin_Vogt06}. 
Modeling global properties of clusters and physical processes 
inside them, such as shocks, fronts, radiobubbles, 
or the heating of the ICM \cite{Peterson_Fabian}, can only be successful 
if the viscosity and thermal conductivity of the ICM are understood 
%(e.g., \cite{Fabian_etal_visc,Dolag_etal_visc,Dennis_Chandran,Ruszkowski_etal}). 
\cite{Fabian_etal_visc}. 
Another fundamental problem is the origin, spatial structure 
and the global dynamical role of the magnetic fields in clusters. 
Turbulent dynamo models again require knowledge 
of the ICM viscosity \cite{SC_dpp05,Ensslin_Vogt06,Subramanian_Shukurov_Haugen},
which itself depends on the field structure, 
so the problem is highly nonlinear and is as yet unsolved.\\
\indent
An additional complication is that in a turbulent plasma, 
pressure anisotropies develop in a spontaneous way 
\cite{SCKHS_brag,SC_dpp05,Sharma_etal}. 
In high-beta plasmas, they trigger a number of instabilities, 
most interestingly, firehose and mirror 
%\cite{Rosenbluth,Chandra_etal,Parker,Vedenov_Sagdeev,Barnes,Hasegawa}. 
\cite{Rosenbluth,Hasegawa}. 
The instabilities are very fast compared to the motions 
of the ICM and give rise to magnetic fluctuations 
at scales as small as $\rho_i$. 
%Even very crude modeling of the effect of these small-scale 
%fluctuations suggests that it can be dramatic: for example, 
%an explosive dynamo may be possible \cite{SC_dpp05}. 
The spatial structure and the saturated amplitude of these 
fluctuations must be understood before quantitative 
models of transport can be constructed. 
In this Letter, we demonstrate
how the nonlinear kinetic theory of these fluctuations can be 
constructed, elucidate the basic physical principle 
behind their nonlinear evolution and 
show that they do not saturate at small quasilinear levels 
\cite{Shapiro_Shevchenko}, but grow nonlinearly to 
large amplitudes ($\dB/B\sim1$). 

{\em The physical origin of pressure anisotropies.}---A fundamental property of a magnetized  
plasma is the conservation of the first 
adiabatic invariant for each particle, $\mu=\vperp^2/2B$
(on time scales $\gg\Omega_i^{-1}$).
%, where $\vperp$ is the particle's velocity perpendicular 
%to the field and $B$ is the field strength. 
This implies that any change in the field strength 
must be accompanied by a corresponding 
change in the perpendicular pressure, $\pperp/B\sim\const$. 
In a heuristic way, we may write \cite{SCKHS_brag}
\bea
\label{pperp_eq}
{1\over\pperp}{d\pperp\over dt} \sim {1\over B}{dB\over dt} - \nuii\,{\pperp-\ppar\over\pperp},
\eea
where the last term represents collisions relaxing 
the pressure anisotropy. 
%(here and in what follows, the subscripts $\perp$ and $\parallel$ 
%refer to the quantities perpendicular and parallel to the 
%magnetic field, respectively). 
On the other hand, 
%since the resistivity of the plasmas 
%such as the ICM is negligible, 
the magnetic field is frozen into the plasma flow velocity $\vu$ 
and the field strength obeys \cite{fn_res}
\bea
\label{B_eq}
{1\over B}{dB\over dt} = \vb\vb:\vdel\vu\sim\go,
\eea
where $d/dt=\dd/\dd t + \vu\cdot\vdel$, 
$\vb=\vB/B$ and $\go$ is the turnover rate of the turbulent motions. 
Taking the two terms in the right-hand side 
of \eqref{pperp_eq} to be comparable and using \eqref{B_eq}, 
we get $\Delta\equiv (\pperp-\ppar)/\pperp \sim \go/\nuii$. 
This is the anisotropy persistently driven by the turbulent motions, 
which are excited at the large (system-size) scales by various macroscopic mechanisms 
%(e.g., \cite{Subramanian_Shukurov_Haugen,Ruszkowski_etal,Chandran_Rasera}). 
\cite{Subramanian_Shukurov_Haugen}.\\ 
\indent
If the turbulence is Kolmogorov, the dominant contribution to 
the turbulent stretching and, therefore, to the pressure anisotropy, 
comes from the viscous scale 
$\lvisc\sim\Re^{-3/4}L$, where $L$ is the outer scale. 
The viscous-scale motions have the characteristic velocity
$u\sim\Re^{-1/4}U$, where $U$ is the characteristic velocity at 
the outer scale. The Reynolds number $\Re=UL/\nu$ is calculated using 
the viscosity of an unmagnetized plasma $\nu\sim\vthi\mfp$
($\vthi$ is the ion thermal speed) 
because for the motions that change the field strength 
the viscosity is not reduced 
by the magnetic field \cite{Braginskii,SC_dpp05}.
% --- we will confirm this at the end of our kinetic calculation below. 
We now introduce a small parameter $\eps\sim \Ma\Re^{-1/4}$, where  
$\Ma=U/\vthi$ is the Mach number \cite{SCKHS_brag}. Then 
%the flow velocity $u$ and its scale $\lvisc$ can be ordered 
we can order $u/\vthi\sim \mfp/\lvisc\sim\eps$, whence  
$\Delta\sim\go/\nuii\sim u\mfp/\lvisc\vthi\sim\eps^2$.
Using again our fiducial parameters for the Hydra A cluster core, 
$U\sim250$~km/s, $\vthi\sim700$~km/s, $L\sim10^{17}$~km 
\cite{Ensslin_Vogt06}, we have $\eps\sim0.1$. 
The relatively small resulting typical anisotropy due to turbulence 
will have a dramatic effect on the magnetic field. 
%Thanks to the existence 
%of a small parameter, we can calculate this effect analytically, 
%but first we give a qualitative preview. 

{\em Qualitative derivation.}---Consider first the firehose instability. 
It is activated when, or in the regions where, $\Delta<0$ 
%\cite{Rosenbluth,Chandra_etal,Parker,Vedenov_Sagdeev}, 
\cite{Rosenbluth}, 
i.e., the magnetic-field strength is decreasing. 
Such events/regions will always exist in a turbulent plasma. 
The growing fluctuations are polarized as Alfv\'en waves, 
with magnetic perturbations perpendicular to the 
original field: $\vB=\vB_0+\dvBperp$. 
Using \eqref{pperp_eq}, we estimate  
\beq
\label{Delta_eq}
\Delta \sim -{|\go|\over\nuii} + 
{\gamma\over\gamma + \nuii}\overline{{\dBperp^2\over B_0^2}},
\eeq
where $\gamma_0=(1/B_0)dB_0/dt<0$, 
the instability growth rate is 
$\gamma=(|\Delta|-2/\beta_i)^{1/2}\kpar\vthi\gg\go$  
(for $k\rho_i\ll1$) 
%\cite{Rosenbluth,Chandra_etal,Parker,Vedenov_Sagdeev,SCKHS_brag}. 
\cite{Rosenbluth,SCKHS_brag}, $\beta_i=4\pi m_in\vthi^2/B_0^2$, 
and the overbar denotes averaging over the fluctuation scales. 
Intuitively, the fluctuations are averaged because 
the particles streaming along the field lines traverse 
the field fluctuations faster than the fluctuations grow
($\kpar\vthi\gg\gamma$). Initially, $\gamma\gg\nuii$;
as $\dBperp$ grows, the instability 
is quenched because the negative anisotropy associated with 
the large-scale turbulence is compensated by a positive anisotropy 
due to the small-scale fluctuations. The amplitude at which the quenching 
occurs is $\dBperp/B_0\sim(|\go|/\nuii)^{1/2}\sim\eps$. 
This estimate can also be obtained 
via a formal quasilinear calculation \cite{Shapiro_Shevchenko}. 
However, it does not, in fact, describe a steady state. 
Indeed, if $\dBperp$ stops changing while 
the unperturbed field $B_0$ continues to decrease, 
%(which it does for a time $\sim|\go|^{-1}$, the decorrelation 
%time of the turbulence; see \eqref{B_eq}), 
the resulting negative pressure anisotropy 
is again uncompensated and the firehose instability will be 
reignited. Since the anisotropy is reduced in the nonlinear 
regime, the growth of the fluctuations 
eventually slows down so that $\gamma\ll\nuii$. Then 
\eqref{Delta_eq} shows that 
the anisotropy stays at the marginal level if 
$(1/B_0^2)d\overline{\dBperp^2}/dt\sim|\go|$, whence
$\dBperp/B_0\sim (|\go|t)^{1/2}$. The physical 
principle of this nonlinear evolution is that the total average 
field strength does not change:
$d\overline{(B_0^2+\dBperp^2)}/dt=0$.\\ 
\indent
Thus, after an initial burst of exponential growth, the firehose 
fluctuations grow secularly until the anisotropy-driving fluid 
motion decorrelates. As this happens on the time scale $\sim|\go|^{-1}$, 
the fluctuations will have time to become large, $\dBperp/B_0\sim1$. 
For Hydra A parameters used above, the time needed for that 
%the emergence of such small-scale ``wrinkled'' magnetic structures 
is $|\go|^{-1}\sim10^6$~yrs. 

\begin{figure}[t]
\centerline{\includegraphics[width=7cm]{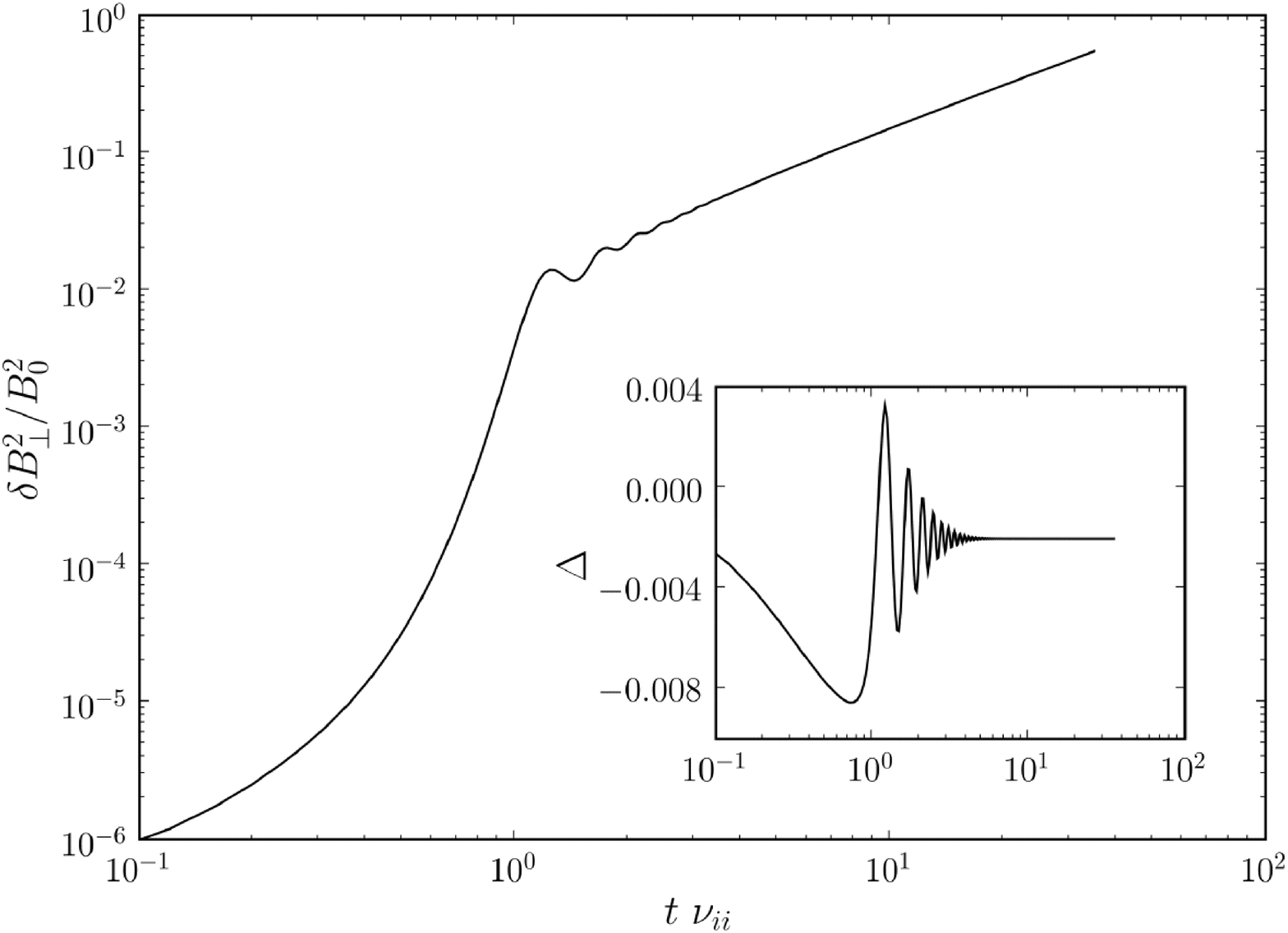}}
\caption{\label{fig_singlek} Evolution of $\dBperp^2(t)/B_0^2$ 
and $\Delta(t)$ (inset) obtained by numerically solving 
\eqsand{dB_eq}{Delta_formula} for a single Fourier mode.
Here $\go/\nuii=0.01$, $\beta_i=1000$ and $\kpar\vthi/\sqrt{2}\,\nuii=10$.}
\vskip-0.5cm
\end{figure}

{\em Kinetic theory.}---We now derive these results 
in a systematic way. Although finite ion Larmor 
radius (FLR) effects are important for the quantitative 
theory of the firehose instability 
%\cite{Yoon_Wu_deAssis,Hellinger_Matsumoto,Horton_etal}, 
\cite{Yoon_Wu_deAssis,Horton_etal}, 
the limit $k\rho_i\ll1$ provides the simplest possible analytical 
framework for elucidating the key elements of the nonlinear physics, which 
persist with FLR \cite{RSC_firehose2}. 
We start with the Kinetic MHD equations \cite{Kulsrud_HPP}, 
valid for $k\rho_i\ll1$ and $\omega\ll\Omega_i$: 
%(frequencies lower than the cyclotron frequency): 
\beq
\label{u_eq}
m_i n\,{d\vu\over dt} = -\vdel\lt(\pperp + {B^2\over8\pi}\rt)
+ \vdel\cdot\lt[\vb\vb\lt(\pperp-\ppar + {B^2\over4\pi}\rt)\rt],
\eeq 
\vskip-0.75cm
\beq
\label{ind_eq}
{d\vB\over dt} = \vB\cdot\vdel\vu.
\eeq
We set $n=\const$ and $\vdel\cdot\vu=0$. 
%(from which we can compute $\pperp$). 
This can be obtained self-consistently, but to reduce 
the amount of formal derivations we simply assume 
incompressibility at all scales %--- a safe choice 
(the motions are subsonic). 
%(the Mach number at the viscous scale is $u/\vthi\sim\eps$). 
The pressure anisotropy is 
$\pperp-\ppar = \int d^3\vv\, m_i(\vperp^2/2 - \vpar^2) f(t,\vx,\vv)$,
where $f$ is the ion distribution function and 
$\vv$ the ion velocity in the frame moving with the mean 
velocity $\vu$. The electron contribution to $\pperp-\ppar$ 
is smaller by $(m_e/m_i)^{1/2}$. 
The ion distribution function satisfies \cite{Kulsrud_HPP}
\bea
\nonumber
{df\over dt}\!\!&+&\!\!\xi v\vb\cdot\vdel f 
- \lt({\dd f\over\dd t}\rt)_\text{c}
= {\vb\cdot\vdel B\over B}\,v\,{1-\xi^2\over2}{\dd f\over \dd\xi}\\
\nonumber
\!\!&+&\!\! \lt(\vb\cdot{d\vu\over dt} - {e\over m_i}\,\Epar\rt)
\lt(\xi{\dd f\over\dd v} + {1-\xi^2\over v}{\dd f\over\dd\xi}\rt)\\
\!\!&-&\!\! \vb\vb:\vdel\vu\lt[{1-3\xi^2\over2}\,v\,{\dd f\over\dd v}
-{3\over2}\lt(1-\xi^2\rt)\xi\,{\dd f\over\dd\xi}\rt],
\label{kin_eq}
\eea
where $v=|\vv|$, $\xi=\vpar/v$ and the last term on the left-hand side is 
the collision operator.\\ 
\indent
We take $\vB=\vB_0+\dvBperp$, $\vu=\vu_0+\dvuperp$, and $\Epar=0$, where 
the {\em slow} fields $\vB_0$, $\vu_0$ (the background turbulence) 
vary at the rate $\go$ on the scale $\lvisc$ of the viscous motions 
(or larger) and the {\em fast} perturbations $\dvBperp$, $\dvuperp$ 
have the growth rate $\gamma$ and wavenumber $k$. 
%Our ability to proceed analytically is due to the existence 
%of the small parameter $\eps$ introduced above. 
We formally order all scales and amplitudes with respect to 
the small parameter $\eps$ introduced above. 
As we see from \eqref{Delta_eq}, it is sensible 
to let the fluctuation growth rate
be (at least) the same order %in the $\eps$ expansion 
as the collision rate: $\gamma\sim\eps k\vthi\sim\nuii$, 
whence $\kpar\sim(\eps\mfp)^{-1}$ \cite{fn_ordering}. 
For the fluid motions, $u_0/\vthi\sim\eps$, $\go\sim\eps^3 k\vthi$, 
and $\lvisc^{-1}\sim\eps^2 k$. The expected fluctuation 
level at which the instability starts being nonlinearly quenched 
tells us to order $\dBperp/B_0\sim\eps$
and, using \eqref{ind_eq}, $\duperp/\vthi\sim\eps^2$.
From \eqref{u_eq} we see that the pressure 
anisotropy is destabilizing only 
if it is not overwhelmed by the magnetic tension, so we 
order $1/\beta_i\sim\Delta\sim\eps^2$.\\ 
\indent
We seek the distribution function $f=f_0 + \df_1 + \df_2 + \cdots$, 
where $f_0$ only has slow variation in space and time. 
To order $\eps$ (the lowest nontrivial order), \eqref{kin_eq} becomes
$\xi v\vb_0\cdot\vdel\df_1 - \lt({\dd f_0/\dd t}\rt)_\text{c} = 0$.
%\beq
%\label{eq_1}
%\xi v\vb\cdot\vdel\df_1 - \lt({\dd f_0\over\dd t}\rt)_\text{c} = 0.
%\eeq
Averaging along the magnetic field, we get 
$(\dd f_0/\dd t)_\text{c}=0$, whence $f_0$ is a Maxwellian: 
%Note that since we are working in the frame moving with 
%the fluid, $f_0$ does not explicitly contain the mean velocity $\vu$, 
%so it is, in fact, a global Maxwellian constant in space and time:
$f_0 = n_0\exp(-v^2/\vthi^2)/(\pi\vthi^2)^{3/2}$. 
Then $\xi v\vb_0\cdot\vdel\df_1 = 0$, i.e., $\df_1$ 
has no fast variation along the magnetic field. 
To order $\eps^2$, 
%\beq
%\label{eq_2}
%{\dd\df_1\over\dd t} + \xi v\vb\cdot\vdel\df_2 
%- \lt({\dd\df_1\over\dd t}\rt)_\text{c} = 0.
%\eeq
we learn, in a similar fashion, that $\df_1$ 
converges to a Maxwellian on the collision time scale 
%(so, discarding density and temperature 
%fluctuations with $\kperp\lvisc\gg1$, $\kpar=0$, which are not unstable, 
(so it can be absorbed into $f_0$) 
and that $\df_2$ has no fast variation along the magnetic field. 
Finally, to order $\eps^3$, 
%\beq
%{\dd\df_2\over\dd t} + \xi v\vb\cdot\vdel\df_3 
%- \lt({\dd\df_2\over\dd t}\rt)_\text{c} = 
%- \vb\vb:\vdel\vu\,{1-3\xi^2\over2}\,v\,{\dd f_0\over\dd v}. 
%\eeq
%Again averaging along the magnetic field and substituting 
%the Maxwellian form of $f_0$, we get  
the kinetic equation averaged along $\vb_0$~is 
\beq
\label{eq_3}
{\dd\df_2\over\dd t} - \lt({\dd\df_2\over\dd t}\rt)_\text{c} = 
\overline{\vb\vb:\vdel\vu}\lt(1-3\xi^2\rt){v^2\over\vthi^2}\,f_0,
\eeq
where the overbar denotes spatial averaging along the field line. 
In order to solve this equation, we assume, as a simple 
model, a pitch-angle-scattering collision operator with 
a constant collision rate: 
$(\dd\df_2/\dd t)_\text{c} = (\nuii/2)(\dd/\dd\xi)(1-\xi^2)\dd\df_2/\dd\xi$. 
While this is not quantitatively correct, it is sufficient 
for our purposes. 
%The solution of \eqref{eq_3} is then  
%$\df_2 = (1-3\xi^2)\int_0^t dt' e^{-3\nuii(t-t')}\overline{\vb\vb:\vdel\vu}(t')
%(v^2/\vthi^2)f_0$. The pressure anisotropy is 
Solving for $\df_2$ and calculating the pressure anisotropy, we find 
\bea
\nonumber
\Delta(t) \equiv {\pperp-\ppar\over p_0} 
= 3\!\int_0^t\!\! dt' e^{-3\nuii(t-t')}\overline{\vb\vb:\vdel\vu}(t')
\qquad\ \ \\
= -{|\go|\over\nuii}\lt(1-e^{-3\nuii t}\rt) + {3\over2}\!\int_0^t\!\! dt' e^{-3\nuii(t-t')}
{d\over dt'}\overline{\dBperp^2(t')\over B_0^2},
\label{Delta_formula}
\eea
where $\go=(1/B_0)dB_0/dt<0$, $p_0=n_0 m_i\vthi^2/2$, and  
we have used \eqref{B_eq}.
\Eqref{Delta_formula} is the quantitative form of \eqref{Delta_eq}. 
Note that it generalizes the Braginskii \cite{Braginskii} formula
$\pperp-\ppar = (p_0/\nuii)\vb\vb:\vdel\vu$, which is only valid 
for fields varying slowly in space in time (cf.\ \cite{SCKHS_brag}).\\
\indent
Applying our ordering to \eqsand{u_eq}{ind_eq}, we get
%arrive, straighforwardly, at the following equation
\beq
{\dd^2\dvBperp\over\dd t^2} = 
{1\over2}\,\vthi^2\lt[\Delta(t) + {2\over\beta_i}\rt]
\dpar^2\dvBperp,
\label{dB_eq}
\eeq
where $\beta_i = 8\pi p_0/B_0^2$ and $\dpar$ is the 
gradient along $\vB_0$.\\ 
%and $\Delta(t)$ is given by \eqref{Delta_formula}.\\  
\indent
\Eqsand{dB_eq}{Delta_formula} describe the evolution of firehose perturbations 
both in the linear and nonlinear regimes. 
%The solution can be represented as a sum of Fourier modes. 
%$\propto e^{\kpar z}$ ($z$ is the coordinate along $\vB_0$). 
Consider the evolution of a single Fourier mode (\figref{fig_singlek}). 
When $\dBperp/B_0\ll\eps$, 
the first (linear) term in \eqref{Delta_formula} dominates and 
the perturbations grow exponentially with the firehose growth rate 
$\gamma = (|\go|/\nuii - 2/\beta_i)^{1/2}\kpar\vthi$. 
Once the nonlinearity becomes significant (at $\dBperp/B_0\sim\eps$), 
the anisotropy is gradually suppressed and, for $t\nuii\gg1$,  
%\bea
%\nonumber
%{\dBperp^2(t)\over B_0^2}\simeq \Lambda t
%- {2\nuii\over\kpar^2\vthi^2 t}
%\qquad\qquad\qquad\qquad\qquad\qquad\quad\ \\
%+\,c_1\sqrt{\Lambda t}\!\int^t\!\!\! dt'
%{e^{-3\nuii t'/2}\over t^{\prime1/4}}\,
%\sin\!\lt(\sqrt{2\Lambda\over3}\kpar\vthi t^{\prime3/2} + c_2\rt),
%\quad
%\eea
${\dBperp^2(t)/B_0^2}\simeq \Lambda t
- {2\nuii/\kpar^2\vthi^2 t}
+ c_1\sqrt{\Lambda t}\int^t dt'
e^{-3\nuii t'/2}t^{\prime-1/4}
\sin\lt(\sqrt{2\Lambda/3}\kpar\vthi t^{\prime3/2} + c_2\rt)$,
where $\Lambda = 2(|\go|-2\nuii/\beta_i)$ and 
$c_1$ and $c_2$ are integration constants. 
The dominant behavior (the first term) 
is the secular growth we already derived qualitatively above. 
The second term is the long-time subdominant correction and 
the third is an oscillatory transient, which decays on 
the collision time scale.\\ 
%The evolution of $\dBperp^2(t)/B_0^2$ and 
%$\Delta(t)$ is shown in \figref{fig_singlek}.\\
\indent
Considering the nonlinear evolution from arbitrary 
initial conditions involving many Fourier modes 
requires inclusion of the FLR terms that set the wave number
of maximum growth. While the spatial structure of the fluctuations 
becomes more complex and a power-law energy spectrum emerges 
\cite{RSC_firehose2}, the key physical 
result derived above persists: the fluctuation energy grows 
secularly with time until finite amplitudes are reached. 

{\em The mirror instability.}---The nonlinear evolution of the mirror 
instability shares some of the features of the firehose, but the full kinetic 
calculation is much more complicated. Here we only present 
a qualitative discussion.\\ 
\indent
The mirror instability is triggered for $\Delta>0$ 
(increasing $B$), has the growth 
rate $\gamma\sim\Delta\,\kpar\vthi$ for $k\rho_i\ll1$, and gives rise to 
growing perturbations of the magnetic-field strength, $\dBpar$
\cite{Hasegawa,SCKHS_brag}. 
The pressure anisotropy is, as before, determined by the changing field 
strength seen on the average by parallel-streaming particles: 
\beq
\label{Delta_mirror}
\Delta \sim {\go\over\nuii} + {1\over\gamma + \nuii}{d\over dt}\overline{\dBpar\over B_0}.
\eeq
For particles traveling the full length of the field line, 
$\overline{\dBpar}=0$; the particles for which $\xi<\xitr\sim|\dBpar/B_0|^{1/2}$ 
are trapped by the fluctuations (``mirrors'') and 
play a key role in the nonlinear dynamics  
%\cite{Kivelson_Southwood,Pantellini}. 
\cite{Kivelson_Southwood}. 
Trapping becomes important when the bounce frequency 
approaches the instability growth rate: 
$\omega_{\rm b}\sim\kpar\vthi\xitr\sim\gamma\sim(\go/\nuii)\kpar\vthi$, 
or $\dBpar/B_0\sim(\go/\nuii)^2\sim\eps^4$. 
For amplitudes above this level, 
$\overline{\dBpar}/B_0\sim\xitr{\dBpar/B_0}\sim-|\dBpar/B_0|^{3/2}$ 
(negative because particles are trapped in the regions of weaker field). 
%Considering the nonlinear stage of the mirror instability, 
We substitute this estimate into \eqref{Delta_mirror},
assume slow evolution ($\gamma\ll\nuii$), and find that 
the marginal state is achieved for 
$\dBpar/B_0\sim (\go t)^{2/3}$. 
This secular growth continues until the 
turbulent motion responsible for the pressure anisotropy decorrelates, 
by which time $\dBpar/B_0\sim1$. The FLR effects, while important 
%\cite{Pokhotelov_etal,Kuznetsov_Passot_Sulem}, 
\cite{Hasegawa,Kuznetsov_Passot_Sulem}, 
are ignored in this qualitative argument, but are unlikely 
to change the main result (secular growth). 

{\em Conclusion.}---We have shown that, 
%owing to the pressure anisotropies persistently generated by turbulence 
in high-beta turbulent plasmas, small-scale magnetic fluctuations are 
continually generated by plasma instabilities and grow nonlinearly 
to large amplitudes, $\dB/B\sim1$, so strongly ``wrinkled'' 
magnetic structures emerge on the fluid time scales. 
%The spatial scale of the wrinkles can be as small as the ion Larmor radius. 
The main difference between our theory and most others 
%existing nonlinear theories for the firehose and mirror fluctuations 
\cite{Shapiro_Shevchenko,Horton_etal,Kivelson_Southwood,Kuznetsov_Passot_Sulem}
is that they consider an {\em initial} 
pressure anisotropy gradually cancelled by fluctuations in 
a collisionless plasma, whereas in our calculation, the anisotropy is 
continually {\em driven} by the turbulent motions and relaxed by 
(weak) collisions; the evolution of the fluctuations is followed 
over times longer than the collision time, up to the fluid time scale. 
The underlying physical principle of the nonlinear evolution 
is the tendency for the growing fluctuations to compensate
on the average the pressure anisotropies 
generated by the turbulence.\\ 
\indent
This mechanism of making small-scale magnetic fields 
is distinct from the fluctuation dynamo, 
which exponentiates the magnetic energy at the turbulent 
stretching rate ($\sim\go$, much slower than the plasma instabilities) 
and produces long filamentary folded structures, so the parallel correlation 
length of the field remains macroscopically large ($\sim$ outer scale) 
\cite{SCTMM_stokes}, in contrast to the instability-produced 
wrinkles with parallel scales possibly as small as the ion gyroscale. 
How the dynamo operates in the presence of the instabilities \cite{SC_dpp05} 
is a subject of an ongoing investigation 
motivated by the fundamental problem of the origin of cosmic magnetism 
in general and of magnetic fields in galaxy clusters in particular.\\ 
\indent
To illustrate the potentially dramatic effect of firehose and mirror fluctuations 
on the transport properties of magnetized turbulent plasmas, consider 
the ICM thermal conduction problem. The standard estimates 
of the electron thermal conductivity in a tangled magnetic field 
are \cite{Rechester_Rosenbluth}
$\kappa_e\sim\vthe\mfp$ if $\mfp\ll\lB$ (collisional), 
$\kappa_e\sim\vthe\mfp\lB/\lRR$ if $\lB\ll\mfp\ll\lRR$ (semicollisional), 
and $\kappa_e\sim\vthe\lB$ if $\mfp\gg\lRR$ (collisionless), 
where $\lB$ is the (parallel) correlation length of the magnetic 
field and $\lRR=\lB\ln(\lB/\rho_e)$ is the Rechester-Rosenbluth length.
In most MHD models \cite{Rechester_Rosenbluth} (including the fluctuation dynamo 
\cite{SCTMM_stokes}), $\lB$ is macroscopic and all three estimates 
yield an effectively isothermal ICM (except at macroscopic scales). 
However, if magnetic wrinkles 
with $\dB/B\sim1$ develop at scales $\sim\rho_i$, 
we have $\lB\sim\rho_i$ and $\lRR\sim\rho_i\ln(\rho_i/\rho_e)\ll\mfp$, 
so $\kappa_e\sim\vthe\rho_i$. For our fiducial Hydra A 
parameters, this is $10^{10}$ times smaller 
than the collisional value, so there is effectively no thermal 
conduction on macroscopic scales. 
The ICM viscosity is similarly reduced, from 
$\vthi\mfp$ to $\vthi\rho_i$ because with $\lB\sim\rho_i$, 
the effective ion mean free path is $\sim\rho_i$.
Curiously, in stronger-field regions 
where $2/\beta_i>\Delta$ and the instabilities are suppressed, 
the transport is more effective: the thermal conductivity and viscosity 
remain large (although highly anisotropic).\\ 
\indent
Due to spatial resolution constraints, the firehose and mirror 
structures are not directly detectable in clusters, but the 
huge changes in the transport coefficients that they may cause 
will have a potentially predictable 
effect on observable large-scale fields and flows \cite{Fabian_etal_visc,Peterson_Fabian}. 
More direct information is available from satellite measurements 
in space plasmas. %, which are in many ways similar to the ICM. 
Mirror structures with $\dB/B\sim1$ have, indeed, been found 
%(e.g., \cite{Lucek_etal,Joy_etal,Sahraoui_etal})
\cite{Lucek_etal} and
there is strong evidence that the directly 
measured temperature anisotropies match the firehose and 
mirror marginal stability conditions
%\cite{Gary_etal_ACE,Kasper_etal,Marsch_Ao_Tu,Hellinger_etal}.  
\cite{Kasper_etal}. 

\begin{acknowledgments} 
This work was supported by UK STFC 
(A.A.S., M.S.R.\ and T.H.), %Newton Trust (T.H.), 
US DOE CMPD %Center for Multiscale Plasma Dynamics 
and the Leverhulme Trust Network for Magnetized Plasma Turbulence. 
\end{acknowledgments}

%\bibliography{sckr_PRL}

\begin{thebibliography}{99}
\expandafter\ifx\csname natexlab\endcsname\relax\def\natexlab#1{#1}\fi
\expandafter\ifx\csname bibnamefont\endcsname\relax
  \def\bibnamefont#1{#1}\fi
\expandafter\ifx\csname bibfnamefont\endcsname\relax
  \def\bibfnamefont#1{#1}\fi
\expandafter\ifx\csname citenamefont\endcsname\relax
  \def\citenamefont#1{#1}\fi
\expandafter\ifx\csname url\endcsname\relax
  \def\url#1{\texttt{#1}}\fi
\expandafter\ifx\csname urlprefix\endcsname\relax\def\urlprefix{URL }\fi
\providecommand{\bibinfo}[2]{#2}
\providecommand{\eprint}[2][]{\url{#2}}

\bibitem{Braginskii}
\bibinfo{author}{\bibfnamefont{S.~I.} \bibnamefont{Braginskii}},
  \bibinfo{journal}{Rev.\ Plasma Phys.} \textbf{\bibinfo{volume}{1}},
  \bibinfo{pages}{205} (\bibinfo{year}{1965}).

\bibitem{SCKHS_brag}
\bibinfo{author}{\bibfnamefont{A.~A.} \bibnamefont{Schekochihin{\ }{et{\ }al.}}},
  \bibinfo{journal}{Astrophys.\ J.} \textbf{\bibinfo{volume}{629}},
  \bibinfo{pages}{139} (\bibinfo{year}{2005}).

\bibitem{SC_dpp05}
\bibinfo{author}{\bibfnamefont{A.~A.} \bibnamefont{Schekochihin}}
  \bibnamefont{and} \bibinfo{author}{\bibfnamefont{S.~C.}
  \bibnamefont{Cowley}}, \bibinfo{journal}{Phys.\ Plasmas}
  \textbf{\bibinfo{volume}{13}}, \bibinfo{pages}{056501}
  (\bibinfo{year}{2006}).

\bibitem{Ensslin_Vogt06}
\bibinfo{author}{\bibfnamefont{T.~A.} \bibnamefont{En{\ss}lin}}
  \bibnamefont{and} \bibinfo{author}{\bibfnamefont{C.}~\bibnamefont{Vogt}},
  \bibinfo{journal}{Astron.\ Astrophys.} \textbf{\bibinfo{volume}{453}},
  \bibinfo{pages}{447} (\bibinfo{year}{2006}).

\bibitem{Peterson_Fabian}
\bibinfo{author}{\bibfnamefont{J.~R.} \bibnamefont{Peterson}} \bibnamefont{and}
  \bibinfo{author}{\bibfnamefont{A.~C.} \bibnamefont{Fabian}},
  \bibinfo{journal}{Phys.\ Rep.} \textbf{\bibinfo{volume}{427}},
  \bibinfo{pages}{1} (\bibinfo{year}{2006});
\bibinfo{author}{\bibfnamefont{M.}~\bibnamefont{Markevitch}} \bibnamefont{and}
  \bibinfo{author}{\bibfnamefont{A.}~\bibnamefont{Vikhlinin}},
  \bibinfo{journal}{Phys.\ Rep.} \textbf{\bibinfo{volume}{443}},
  \bibinfo{pages}{1} (\bibinfo{year}{2007}).

\bibitem{Fabian_etal_visc}
\bibinfo{author}{\bibfnamefont{A.}~\bibnamefont{Fabian{\ }{et{\ }al.}}},
  \bibinfo{journal}{Mon.\ Not.\ R.\ Astron.\ Soc.}
  \textbf{\bibinfo{volume}{363}}, \bibinfo{pages}{891} (\bibinfo{year}{2005});
\bibinfo{author}{\bibfnamefont{K.}~\bibnamefont{Dolag{\ }{et{\ }al.}}},
  \bibinfo{journal}{Mon.\ Not.\ R.\ Astron.\ Soc.}
  \textbf{\bibinfo{volume}{364}}, \bibinfo{pages}{753} (\bibinfo{year}{2005});
\bibinfo{author}{\bibfnamefont{T.~J.} \bibnamefont{Dennis}} \bibnamefont{and}
  \bibinfo{author}{\bibfnamefont{B.~D.~G.} \bibnamefont{Chandran}},
  \bibinfo{journal}{Astrophys.\ J.} \textbf{\bibinfo{volume}{622}},
  \bibinfo{pages}{205} (\bibinfo{year}{2005});
\bibinfo{author}{\bibfnamefont{M.}~\bibnamefont{Ruszkowski{\ }{et{\ }al.}}},
  \bibinfo{journal}{Mon.\ Not.\ R.\ Astron.\ Soc.} \textbf{\bibinfo{volume}{378}},
  \bibinfo{pages}{662} (\bibinfo{year}{2007}).

\bibitem{Subramanian_Shukurov_Haugen}
\bibinfo{author}{\bibfnamefont{K.}~\bibnamefont{Subramanian{\ }{et{\ }al.}}},
%  \bibinfo{author}{\bibfnamefont{A.}~\bibnamefont{Shukurov}}, \bibnamefont{and}
%  \bibinfo{author}{\bibfnamefont{N.~E.~L.} \bibnamefont{Haugen}},
  \bibinfo{journal}{Mon.\ Not.\ R.\ Astron.\ Soc.}
  \textbf{\bibinfo{volume}{366}}, \bibinfo{pages}{1437} (\bibinfo{year}{2006}).

\bibitem{Sharma_etal}
\bibinfo{author}{\bibfnamefont{P.}~\bibnamefont{Sharma{\ }{et{\ }al.}}},
  \bibinfo{journal}{Astrophys.\ J.} \textbf{\bibinfo{volume}{637}},
  \bibinfo{pages}{952} (\bibinfo{year}{2006}).

\bibitem{Rosenbluth}
\bibinfo{author}{\bibfnamefont{M.~N.} \bibnamefont{Rosenbluth}},
  \bibinfo{journal}{LANL Report LA-2030}
  (\bibinfo{year}{1956});
\bibinfo{author}{\bibfnamefont{S.}~\bibnamefont{Chandrasekhar{\ }{et{\ }al.}}},
%  \bibinfo{author}{\bibfnamefont{A.~N.} \bibnamefont{Kaufman}},
%  \bibnamefont{and} \bibinfo{author}{\bibfnamefont{K.~M.} \bibnamefont{Watson}}, 
  \bibinfo{journal}{Proc.\ R.\ Soc.\ London~A}
  \textbf{\bibinfo{volume}{245}}, \bibinfo{pages}{435} (\bibinfo{year}{1958});
\bibinfo{author}{\bibfnamefont{E.~N.} \bibnamefont{Parker}},
  \bibinfo{journal}{Phys.\ Rev.} \textbf{\bibinfo{volume}{109}},
  \bibinfo{pages}{1874} (\bibinfo{year}{1958});
\bibinfo{author}{\bibfnamefont{A.~A.} \bibnamefont{Vedenov}} \bibnamefont{and}
  \bibinfo{author}{\bibfnamefont{R.~Z.} \bibnamefont{Sagdeev}},
  \bibinfo{journal}{Sov.\ Phys.\ Dokl.} \textbf{\bibinfo{volume}{3}},
  \bibinfo{pages}{278} (\bibinfo{year}{1958});
\bibinfo{author}{\bibfnamefont{A.}~\bibnamefont{Barnes}},
  \bibinfo{journal}{Phys.\ Fluids} \textbf{\bibinfo{volume}{9}},
  \bibinfo{pages}{1483} (\bibinfo{year}{1966}).

\bibitem{Hasegawa}
\bibinfo{author}{\bibfnamefont{A.}~\bibnamefont{Hasegawa}},
  \bibinfo{journal}{Phys.\ Fluids} \textbf{\bibinfo{volume}{12}},
  \bibinfo{pages}{2642} (\bibinfo{year}{1969}).

\bibitem{Shapiro_Shevchenko}
\bibinfo{author}{\bibfnamefont{V.~D.} \bibnamefont{Shapiro}} \bibnamefont{and}
  \bibinfo{author}{\bibfnamefont{V.~I.} \bibnamefont{Shevchenko}},
  \bibinfo{journal}{Sov.\ Phys.\ JETP} \textbf{\bibinfo{volume}{18}},
  \bibinfo{pages}{1109} (\bibinfo{year}{1964}).

\bibitem{fn_res} We consider magnetic 
fields at scales greater than $\rho_i$. In plasmas such as the ICM, 
the resistive scale is smaller than $\rho_i$ \cite{SC_dpp05}, 
so we neglect the resistivity (this
hinges on assuming that the characteristic scale of the magnetic field 
is {\em not} determined by the resistive cutoff, unlike in 
standard, collisionally dominated MHD \cite{SCTMM_stokes}). 
We also assume $\vdel\cdot\vu=0$ because the plasma motions are subsonic.

\bibitem{SCTMM_stokes}
\bibinfo{author}{\bibfnamefont{A.~A.} \bibnamefont{Schekochihin{\ }{et{\ }al.}}},
  \bibinfo{journal}{Astrophys.\ J.} \textbf{\bibinfo{volume}{612}},
  \bibinfo{pages}{276} (\bibinfo{year}{2004}).

%\bibitem{Chandran_Rasera}
%\bibinfo{author}{\bibfnamefont{B.~D.~G.} \bibnamefont{Chandran}}
%  \bibnamefont{and} \bibinfo{author}{\bibfnamefont{Y.}~\bibnamefont{Rasera}},
%  \bibinfo{journal}{Astrophys.\ J.}  (\bibinfo{year}{2007}), \bibinfo{note}{in
%  press (arXiv:0707.2351)}.

\bibitem{Yoon_Wu_deAssis}
\bibinfo{author}{\bibfnamefont{P.~H.} \bibnamefont{Yoon{\ }{et{\ }al.}}},
%  \bibinfo{author}{\bibfnamefont{C.~S.} \bibnamefont{Wu}}, \bibnamefont{and}
%  \bibinfo{author}{\bibfnamefont{A.~S.} \bibnamefont{de{\ }Assis}},
  \bibinfo{journal}{Phys.\ Fluids B} \textbf{\bibinfo{volume}{5}},
  \bibinfo{pages}{1971} (\bibinfo{year}{1993});
\bibinfo{author}{\bibfnamefont{P.}~\bibnamefont{Hellinger}} \bibnamefont{and}
  \bibinfo{author}{\bibfnamefont{H.}~\bibnamefont{Matsumoto}},
  \bibinfo{journal}{J.\ Geophys.\ Res.} \textbf{\bibinfo{volume}{105}},
  \bibinfo{pages}{10519} (\bibinfo{year}{2000}).

\bibitem{Horton_etal}
\bibinfo{author}{\bibfnamefont{W.}~\bibnamefont{Horton{\ }{et{\ }al.}}},
  \bibinfo{journal}{J.\ Geophys.\ Res.} \textbf{\bibinfo{volume}{109}},
  \bibinfo{pages}{A09216} (\bibinfo{year}{2004}).

\bibitem{RSC_firehose2}
\bibinfo{author}{\bibfnamefont{M.~S.} \bibnamefont{Rosin{\ }{et{\ }al.}}},
%  \bibinfo{author}{\bibfnamefont{A.~A.} \bibnamefont{Schekochihin}},
%  \bibnamefont{and} \bibinfo{author}{\bibfnamefont{S.~C.} \bibnamefont{Cowley}}
  (\bibinfo{year}{2007}), \bibinfo{note}{in preparation}.

\bibitem{Kulsrud_HPP}
\bibinfo{author}{\bibfnamefont{R.~M.} \bibnamefont{Kulsrud}}, in
  \emph{\bibinfo{booktitle}{Handbook of Plasma Physics, Vol.~1}}, edited by
  \bibinfo{editor}{\bibfnamefont{A.~A.} \bibnamefont{Galeev}} \bibnamefont{and}
  \bibinfo{editor}{\bibfnamefont{R.~N.} \bibnamefont{Sudan}}
  (\bibinfo{publisher}{North-Holland}, \bibinfo{address}{Amsterdam},
  \bibinfo{year}{1983}), p. \bibinfo{pages}{115}.

\bibitem{fn_ordering} This ordering is a formal 
imposition that will help us proceed in a systematic fashion. In reality, 
the most linearly unstable wave number is determined by the FLR  
modification of the growth rate \cite{Yoon_Wu_deAssis}. 
%We also order $\kperp\sim\kpar$, which is indeed correct 
%for the most unstable modes \cite{Yoon_Wu_deAssis}. 

\bibitem{Kivelson_Southwood}
\bibinfo{author}{\bibfnamefont{M.~G.} \bibnamefont{Kivelson}} \bibnamefont{and}
  \bibinfo{author}{\bibfnamefont{D.~S.} \bibnamefont{Southwood}},
  \bibinfo{journal}{J.\ Geophys.\ Res.} \textbf{\bibinfo{volume}{101}},
  \bibinfo{pages}{17365} (\bibinfo{year}{1996}).
%\bibinfo{author}{\bibfnamefont{P.~G.~E.} \bibnamefont{Pantellini}},
%  \bibinfo{journal}{J.\ Geophys.\ Res.} \textbf{\bibinfo{volume}{103}},
%  \bibinfo{pages}{4789} (\bibinfo{year}{1998}).

\bibitem{Kuznetsov_Passot_Sulem}
%\bibinfo{author}{\bibfnamefont{O.~A.} \bibnamefont{Pokhotelov{\ }{et{\ }al.}}},
%  \bibinfo{journal}{J.\ Geophys.\ Res.} \textbf{\bibinfo{volume}{109}},
%  \bibinfo{pages}{A09213} (\bibinfo{year}{2004});
\bibinfo{author}{\bibfnamefont{E.~A.} \bibnamefont{Kuznetsov{\ }{et{\ }al.}}},
%  \bibinfo{author}{\bibfnamefont{T.}~\bibnamefont{Passot}}, \bibnamefont{and}
%  \bibinfo{author}{\bibfnamefont{P.~L.} \bibnamefont{Sulem}},
  \bibinfo{journal}{Phys.\ Rev.\ Lett.} \textbf{\bibinfo{volume}{98}},
  \bibinfo{pages}{235003} (\bibinfo{year}{2007});
\bibinfo{author}{\bibfnamefont{P.} \bibnamefont{Hellinger}},
  \bibinfo{journal}{Phys.\ Plasmas} \textbf{\bibinfo{volume}{14}},
  \bibinfo{pages}{082105} (\bibinfo{year}{2007}).

\bibitem{Rechester_Rosenbluth}
\bibinfo{author}{\bibfnamefont{A.~B.} \bibnamefont{Rechester}} 
  \bibnamefont{and}
  \bibinfo{author}{\bibfnamefont{M.~N.} \bibnamefont{Rosenbluth}},
  \bibinfo{journal}{Phys.\ Rev.\ Lett.} \textbf{\bibinfo{volume}{40}},
  \bibinfo{pages}{38} (\bibinfo{year}{1978});
\bibinfo{author}{\bibfnamefont{B.~D.~G.} \bibnamefont{Chandran}} 
  \bibnamefont{and}
  \bibinfo{author}{\bibfnamefont{S.~C.} \bibnamefont{Cowley}},
  \bibinfo{journal}{Phys.\ Rev.\ Lett.} \textbf{\bibinfo{volume}{80}},
  \bibinfo{pages}{3077} (\bibinfo{year}{1998}).
%\bibinfo{author}{\bibfnamefont{R.} \bibnamefont{Narayan}} 
%  \bibnamefont{and}
%  \bibinfo{author}{\bibfnamefont{M.} \bibnamefont{Medvedev}},
%  \bibinfo{journal}{Astrophys.~J.} \textbf{\bibinfo{volume}{562}},
%  \bibinfo{pages}{L129} (\bibinfo{year}{2001}).
%\bibinfo{author}{\bibfnamefont{B.~D.~G.} \bibnamefont{Chandran}} 
%  \bibnamefont{and}
%  \bibinfo{author}{\bibfnamefont{J.~L.} \bibnamefont{Maron}},
%  \bibinfo{journal}{Astrophys.~J.} \textbf{\bibinfo{volume}{602}},
%  \bibinfo{pages}{170} (\bibinfo{year}{2004}).

\bibitem{Lucek_etal}
\bibinfo{author}{\bibfnamefont{E.~A.} \bibnamefont{Lucek{\ }{et{\ }al.}}},
  \bibinfo{journal}{Ann.\ Geophys.} \textbf{\bibinfo{volume}{19}},
  \bibinfo{pages}{1421} (\bibinfo{year}{2001});
%\bibinfo{author}{\bibfnamefont{S.~P.} \bibnamefont{Joy{\ }{et{\ }al.}}},
%  \bibinfo{journal}{J.\ Geophys.\ Res.} \textbf{\bibinfo{volume}{111}},
%  \bibinfo{pages}{A12212} (\bibinfo{year}{2006});
\bibinfo{author}{\bibfnamefont{F.}~\bibnamefont{Sahraoui{\ }{et{\ }al.}}},
  \bibinfo{journal}{Phys.\ Rev.\ Lett.} \textbf{\bibinfo{volume}{96}},
  \bibinfo{pages}{075002} (\bibinfo{year}{2006}).

\bibitem{Kasper_etal}
%\bibinfo{author}{\bibfnamefont{S.~P.} \bibnamefont{Gary{\ }{et{\ }al.}}},
%  \bibinfo{journal}{Geophys. Res. Lett.} \textbf{\bibinfo{volume}{28}},
%  \bibinfo{pages}{2759} (\bibinfo{year}{2001});
%\bibinfo{author}{\bibfnamefont{J.~C.} \bibnamefont{Kasper{\ }{et{\ }al.}}},
%  \bibinfo{author}{\bibfnamefont{A.~J.} \bibnamefont{Lazarus}},
%  \bibnamefont{and} \bibinfo{author}{\bibfnamefont{S.~P.} \bibnamefont{Gary}},
%  \bibinfo{journal}{Geophys. Res. Lett.} \textbf{\bibinfo{volume}{29}},
%  \bibinfo{pages}{1839} (\bibinfo{year}{2002});
%\bibinfo{author}{\bibfnamefont{E.} \bibnamefont{Marsch{\ }{et{\ }al.}}},
%  \bibinfo{author}{\bibfnamefont{X.-Z.} \bibnamefont{Ao}}, \bibnamefont{and}
%  \bibinfo{author}{\bibfnamefont{C.-Y.} \bibnamefont{Tu}},
%  \bibinfo{journal}{J.\ Geophys.\ Res.} \textbf{\bibinfo{volume}{109}},
%  \bibinfo{pages}{A04102} (\bibinfo{year}{2004});
\bibinfo{author}{\bibfnamefont{P.}~\bibnamefont{Hellinger{\ }{et{\ }al.}}},
  \bibinfo{journal}{Geophys. Res. Lett.} \textbf{\bibinfo{volume}{33}},
  \bibinfo{pages}{L09101} (\bibinfo{year}{2006}).

\end{thebibliography}
%\end{document}

\end{document}